\documentclass[twocolumn]{aastex63}
\usepackage{gensymb}
\usepackage{autobreak}
\usepackage{threeparttable}
\submitjournal{AJ}

\shorttitle{Palomar 5}
\shortauthors{Xu et al.}

\begin{document}

\title{New Determination of Fundamental Properties of Palomar 5 Using Deep DESI Imaging Data}

\correspondingauthor{Hu Zou}
\email{zouhu@nao.cas.cn}

\author{Xin Xu}
\affiliation{Key Laboratory of Optical Astronomy, National Astronomical Observatories, Chinese Academy of Sciences, Beijing 100012, People’s Republic of China}
\affiliation{School of Astronomy and Space Science, University of Chinese Academy of Sciences, Beijing 101408, People’s Republic of China}

\author{Hu Zou}
\affiliation{Key Laboratory of Optical Astronomy, National Astronomical Observatories, Chinese Academy of Sciences, Beijing 100012, People’s Republic of China}

\author{Xu Zhou}
\affiliation{Key Laboratory of Optical Astronomy, National Astronomical Observatories, Chinese Academy of Sciences, Beijing 100012, People’s Republic of China}

\author{Jundan Nie}
\affiliation{Key Laboratory of Optical Astronomy, National Astronomical Observatories, Chinese Academy of Sciences, Beijing 100012, People’s Republic of China}

\author{Zhimin Zhou}
\affiliation{Key Laboratory of Optical Astronomy, National Astronomical Observatories, Chinese Academy of Sciences, Beijing 100012, People’s Republic of China}

\author{Jun Ma}
\affiliation{Key Laboratory of Optical Astronomy, National Astronomical Observatories, Chinese Academy of Sciences, Beijing 100012, People’s Republic of China}
\affiliation{School of Astronomy and Space Science, University of Chinese Academy of Sciences, Beijing 101408, People’s Republic of China}

\author{Tianmeng Zhang}
\affiliation{Key Laboratory of Optical Astronomy, National Astronomical Observatories, Chinese Academy of Sciences, Beijing 100012, People’s Republic of China}

\author{Jiali Wang}
\affiliation{Key Laboratory of Optical Astronomy, National Astronomical Observatories, Chinese Academy of Sciences, Beijing 100012, People’s Republic of China}

\author{Suijian Xue}
\affiliation{Key Laboratory of Optical Astronomy, National Astronomical Observatories, Chinese Academy of Sciences, Beijing 100012, People’s Republic of China}



\begin{abstract}

The legacy imaging surveys for the Dark Energy Spectroscopic Instrument project provides multiple-color photometric data, which are about 2 mag deeper than the SDSS. In this study, we redetermine the fundamental properties for an old halo globular cluster of Palomar 5 based on these new imaging data, including structure parameters, stellar population parameters, and luminosity and mass functions. These characteristics, together with its tidal tails, are key for dynamical studies of the cluster and constraining the mass model of the Milky Way. By fitting the King model to the radial surface density profile of Palomar 5, we derive the core radius of $r_c=2.^{\prime}96\pm0.^{\prime}11$, tidal radius of $r_t=17.^{\prime}99\pm1.^{\prime}49$, and concentration parameter  of  $c=0.78\pm0.04$. We apply a Bayesian analysis method to derive the stellar population properties and get an age of $11.508\pm0.027$ Gyr, metallicity of [Fe/H]$=-1.798\pm0.014$, reddening of $E(B-V) = 0.0552\pm0.0005$, and distance modulus of $(m-M)_0=16.835\pm0.006$. The main-sequence luminosity and mass functions for both the cluster center, and tidal tails are investigated. The luminosity and mass functions at different distances from the cluster center suggest that there is obvious spatial mass segregation. Many faint low-mass stars have been evaporated at the cluster center and the tidal tails are enhanced by low-mass stars. Both the concentration and relaxation times suggest that Palomar 5 is a totally relaxed system.

\end{abstract}

\keywords{globular clusters: individual (Palomar 5) --- 
stars: fundamental parameters --- stars: luminosity function}


\section{Introduction} \label{sec:intro}

Globular clusters (GCs) representing the oldest objects in the universe have witnessed the early formation history of the Galaxy. Most GCs are distributed in the Milky Way halo. GCs were once regarded as simple stellar populations due to similar initial birth conditions of their star members \citep{1986ASSL..122..195R}. However, recent studies show that nearly all the GCs can be recognized as multiple stellar populations \citep{2015AJ....149...91P, 2017MNRAS.464.3636M, 2018ARA&A..56...83B}. The stellar population parameters of GCs can be determined accurately by comparing the observed data with theoretical models. GCs also provide an ideal laboratory to study star formation and evolution, stellar physics, and chemical enrichment processes \citep{2018ApJ...865..160M}. They have long been considered as the probes of early formation history and evolutionary process of the Galaxy, offering important information for understanding the structure of the Galactic halo.

GCs lose their member stars by both internal dynamical processes, including stellar evolution, binary heating, and two-body relaxation, and external dynamical processes induced by the Galactic gravitational field \citep{2003AJ....126.2385O,2010MNRAS.401..131S}. Some GCs with extratidal structures show the departure from a classic tidally limited profile \citep{2015MNRAS.446.3297K}. The stars escaping the cluster through evaporation and tidal stripping can form thin and kinematically cold tidal tails \citep{1999A&A...352..149C,2003MNRAS.340..227B}. The existence of tidal tails indicates the continuous mass loss of member stars, which may eventually result in the complete dissolution of the cluster \citep{2015MNRAS.446.3297K}. The tidal tails are usually aligned with the orbit of the GC, providing observational constraints on the gravitational potential and dark matter halo of the Milky Way \citep{2011MNRAS.411.1989Z,2012A&A...546L...7M,2015ApJ...811..123F,2015ApJ...799...28P}. 

Among all Galactic GCs, Palomar 5 stands out due to its unusual properties and visible long tidal tails. It is a remote halo GC located at a distance of 23.2 kpc from the sun \citep{1996AJ....112.1487H}. Palomar 5 is faint and sparse and has low mass, luminosity, velocity dispersion, and central concentration and has a highly eccentric orbit \citep{2002AJ....124.1497O,2015MNRAS.446.3297K}. These peculiar characteristics indicate that Palomar 5 has lost most of its mass into tidal tails. Many works in literature focus on structure parameters and stellar population properties of this cluster. We summarize these parameters in Tables \ref{tab:stlit} and \ref{tab:plit}. Over the last decades, many efforts have also been made to investigate the tidal tails of Palomar 5. The tidal tails were first detected by \citet{2001ApJ...548L.165O} and revisited by many other works \citep{2003AJ....126.2385O, 2006ApJ...641L..37G, 2009RAA.....9.1131Z, 2012ApJ...760...75C, 2020ApJ...889...70B}. The simulation of \citet{2004AJ....127.2753D} showed that Palomar 5 will be destroyed at its next disk passage.  

Based on the photometric data from the Dark Energy Spectroscopic  Instrument (DESI) legacy imaging surveys \citep{2019AJ....157..168D}, we intend to carry out a series of studies of Palomar 5. The latest released data
\footnote{\url{http://legacysurvey.org/}} 
include multicolor photometry for stars and galaxies covering a sky area of about 20,000 deg$^2$. The imaging data are about 2 mag deeper than the SDSS, which can be used to better determine the cluster properties. The stellar population parameters can be derived from fitting the observed color-magnitude diagrams (CMDs) to theoretical isochrones. These basic parameters are of key importance to determine membership probabilities and detect member stars \citep{2015AJ....150...61W} and investigate the relation between mass and absolute magnitude for different stellar models \citep{2001AJ....122.3231G}, etc. The structural parameters of Palomar 5 can be determined by the King model fitting, which are usually used to check the existence of tidal tails for star clusters \citep{1995AJ....109.2553G, 2001ApJ...548L.165O}. These parameters can also be used to derive the dynamical mass together with the velocity dispersion \citep{1966AJ.....71...64K,2002AJ....124.1497O}. The luminosity functions reflecting the mass distributions can be used to explore the mass segregation \citep{2004AJ....128.2274K}. Combining the SDSS and DESI imaging data, we can obtain an observational baseline of about 10--15 yr, which can yield good measurements of proper motions for stars down to the SDSS limiting magnitude. With proper motions and multicolor photometry, we will also explore in detail the tidal tails of Palomar 5 and its dynamical status, statistically study Galactic star clusters, discover more substructures in the Galactic halo, and further constrain the gravitational potential and mass compositions of the Milky Way.

As the first article in a series, this paper focuses on the new determination of the fundamental properties of Palomar 5 based on the deep DESI imaging data. The paper is organized as follows. In Section \ref{sec:data}, we give an introduction of the photometric data. Section \ref{sec:property} presents the derivations of the basic parameters of Palomar 5 as well as comparisons with the results from literature. In this section, we obtain the structure parameters by fitting the radial surface density profile with the King model and the stellar population parameters by fitting the multiwavelength photometric data with theoretical isochrones. In Section \ref{sec:luminosity}, we analyze the luminosity and mass functions of the cluster and its tidal tails and present the mass segregation and mass loss. Section \ref{sec:summary} gives a summary.

\section{Photometric Data} \label{sec:data}

The DESI project is a spectroscopic survey program, which will measure the effect of dark energy on the expansion of the universe by using 5000-fiber multiobject spectrographs installed on the Mayall 4 m telescope \citep{2016arXiv161100036D}. The DESI legacy imaging surveys provide $g$-, $r$-, and $z$-band optical images to select spectroscopic targets for DESI \citep{2019AJ....157..168D}.  They consist of three different components: the Beijing-Arizona Sky Survey \citep[BASS;][]{2017PASP..129f4101Z}, the Dark Energy Camera Legacy Survey \citep[DECaLS;][]{2016AAS...22831701B}, and the Mayall $z$-band Legacy Survey \citep[MzLS;][]{2016AAS...22831702S}.  These surveys cover a total area of about 14,000 deg$^2$ and the fiducial depths at $5\sigma$ are $g=24.0$, $r=23.4$ and $z=22.5$ mag. These depths are set to meet the requirements for DESI target selections.  

\begin{figure*}[ht!]
\centering
\includegraphics[width=0.8\textwidth]{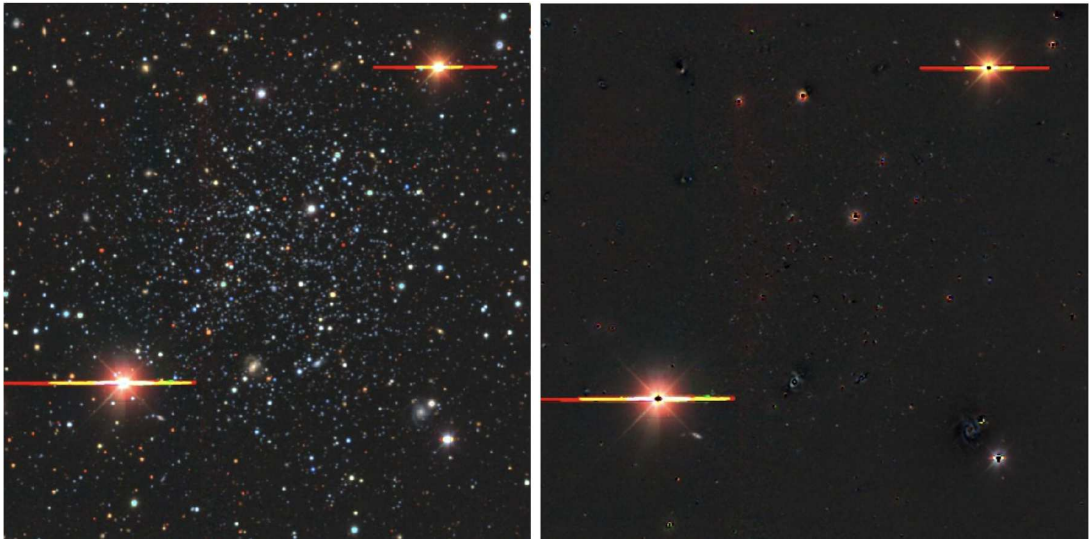}
\caption{Left:  composite color image of Palomar 5 (blue channel: $g$ band, green channel: $r$ band, and red channel: $z$ band). Right: the residual image generated by subtracting object models from the original image. North is up and east is left.  \label{fig:image}}
\end{figure*}

The legacy surveys are considered as public projects, whose raw data are made available immediately after they are taken. The processed images and catalogs are released almost twice a year. The latest data release is DR8,\footnote{\url{https://www.legacysurvey.org/dr8/}} which includes the data from all three surveys as well as the data taken by the Dark Energy Survey \citep[DES;][]{2016MNRAS.460.1270D}. The DES increases the sky coverage in the south Galactic cap to lower declinations. The final sky area reaches about 20,000 deg$^2$.  The source catalogs in DR8 are constructed by The Tractor,\footnote{\url{https://github.com/dstndstn/tractor}} which models each source with parametric profiles, including delta function for point source,  de Vaucouleurs law, exponential law, and composite profile of de Vaucouleurs plus exponential. The best-fit model convolved with specific PSF yields six morphological types: point sources (PSF), round exponential galaxies, de Vaucouleurs profiles (DEV), normal exponential profiles (EXP), and composite profiles (COMP).  Palomar 5 is located in the footprint of the DECaLS. The DECaLS imaging data are somewhat deeper than the fiducial depth requirements (about 0.5 mag deeper). Figure \ref{fig:image} shows a composite color image of Palomar 5 taken from the sky viewer in the official website\footnote{\url{https://www.legacysurvey.org/viewer}}. The residual image, which is generated by subtracting the source models from the observed image, is also presented in this figure. The residual image looks quite clean for unsaturated stars, indicating the good performance of source detection and profile modeling. 

\subsection{Photometric depths}
\begin{figure*}[ht!]
\centering
\includegraphics[width=1\textwidth]{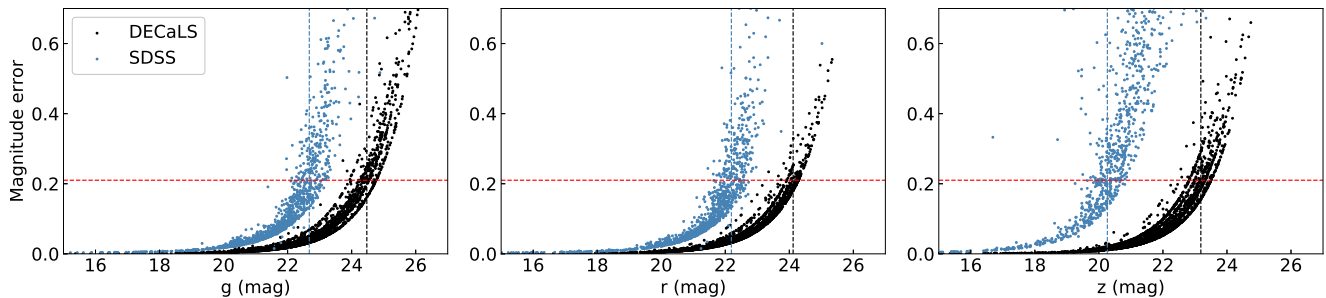}
\caption{Magnitude error as function of magnitude for $g$ (left), $r$ (middle), and $z$ (right) bands. The black and  blue points represent the DECaLS and SDSS data, respectively. The horizontal line presents the magnitude error of 0.21 mag (i.e. S/N=5). The black and blue vertical dashed line shows the $5\sigma$ magnitude limits for the DECaLS and SDSS, respectively.  \label{fig:magerr}}
\end{figure*}

We obtain the photometric data of point sources (classified as ``PSF") in an area of about 3\degree$\times$3\degree around Palomar 5. The magnitudes are corrected with the foreground Galactic extinction based on the \citet{1998ApJ...500..525S} dust reddening map.  Hereafter, all magnitudes used in this paper are PSF magnitudes after foreground extinction correction. In order to show the actual imaging depths in the region of Palomar 5, we present the diagram of magnitude error as function of magnitude in Figure \ref{fig:magerr}. The DECaLS magnitude limits at S/N $=$ 5 (at the magnitude error of $\sigma = 0.21$ mag) are $g$ = 24.47, $r$ = 24.12, and $z$ = 23.18. Most of previous studies about Palomar 5 were based on the SDSS data, so we also compare the SDSS magnitude limits in Figure \ref{fig:magerr}. The photometric systems between SDSS and DECaLS are different. \citet{2019AJ....157..168D} presented a series of system transformation equations for transforming the SDSS magnitude to the DECaLS magnitude. We apply these transformation equations to the SDSS PSF magnitude, and again, the magnitudes are corrected for the foreground Galactic extinction. The $5\sigma$ limiting magnitudes are $g$ = 22.68, $r$ = 22.20, and $z$=20.27.  Generally, the DECaLS $g$ and $r$ bands are nearly 2 mag deeper than the SDSS data, while the DECaLS $z$ band is almost 3 mag deeper.  

\subsection{Completeness of source detection}
Because only stars are considered in this paper, we select point sources identified as ``PSF" in the DECaLS catalog. In the rest of the paper, we only consider those stars with $r < 23.2$ mag (error of about 0.1 mag) to eliminate the objects with large astrometric and photometric uncertainties. We assess the completeness of the star detection for the DECaLS catalog. Artificial star tests are made in order to determine the completeness functions: artificial stars are first added to the observed images, and they are then reidentified by using the same data reduction pipeline. Because the DECaLS photometric catalog is adopted in this paper and it is hard to reproduce the corresponding pipeline, we independently detect sources by using SExtractor \citep{1996A&AS..117..393B}  and try to make source detection as consistent as possible.

\begin{figure}[ht!]
\centering
\includegraphics[width=\columnwidth]{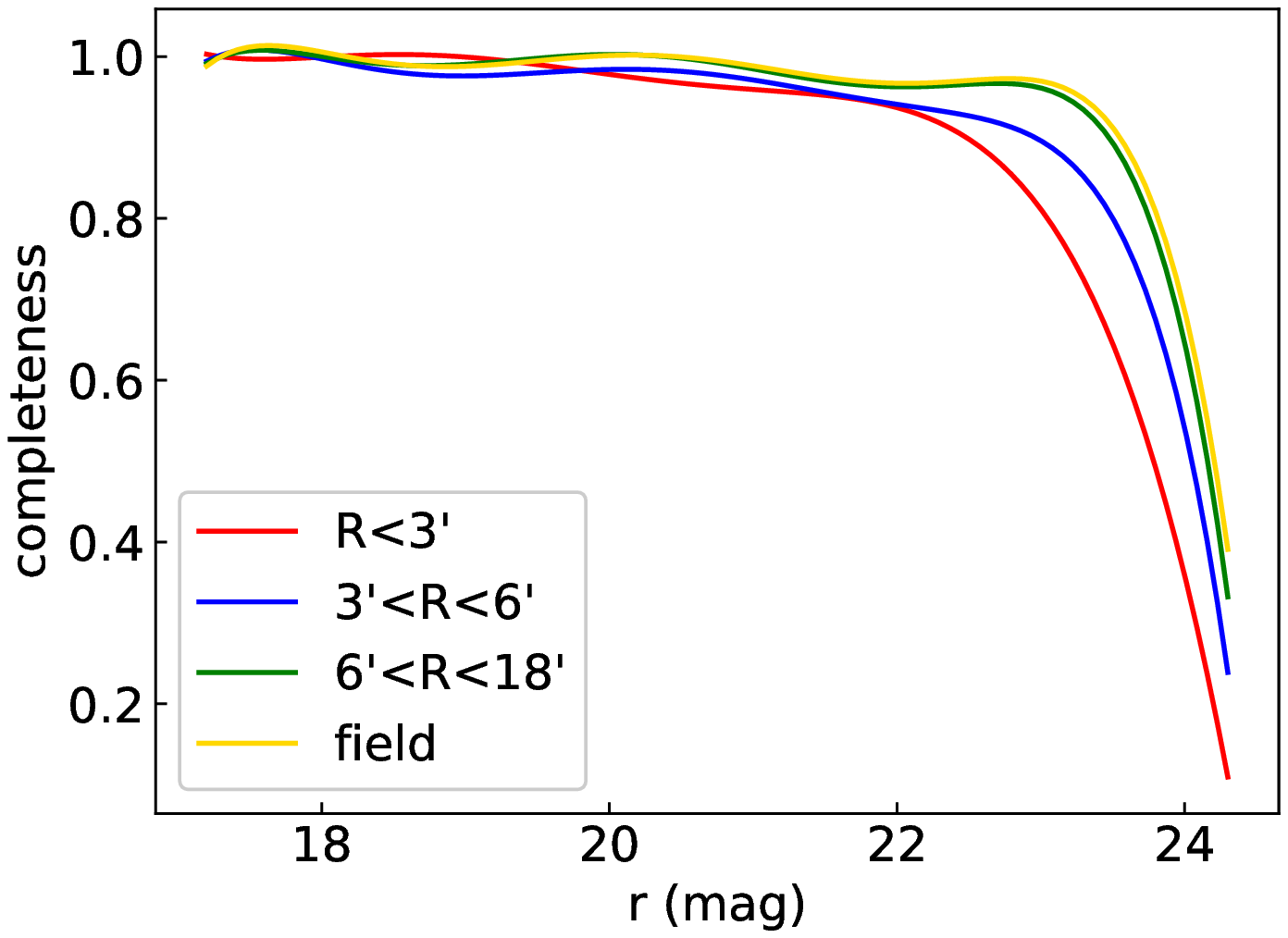}
\caption{The completeness for stars as a function of $r$-band magnitude and distance from the cluster center. The solid lines are the best-fit, sixth-order polynomial curves. The red, blue, and green lines represent the completeness at $R<3\arcmin$, $3\arcmin<R<6\arcmin$, and $6\arcmin<R<18\arcmin$, respectively. The yellow line shows the completeness of field stars.  \label{fig:completeness}}
\end{figure}

The DECaLS $r$-band images are obtained. For each observed image, we add 3000 artificial stars with random positions and magnitudes (ranging from 15 to 25 mag).  All sources are detected by SExtractor and  cross-matched with input artificial stars. The above process is repeated 10 times and then the average completeness is calculated. To reflect the crowding effect, we perform the completeness measurements in three cluster regions ($R<3{\arcmin}$, $3\arcmin<R<6\arcmin$, and $6\arcmin<R<18\arcmin$) and one field-star region. Figure \ref{fig:completeness} shows the completeness as function of $r$-band magnitude and distance from the cluster center. We get a smooth completeness curve by fitting a sixth-order polynomial to the data. The crowding effect is shown clearly in this figure. The completeness becomes worse when it is closer to the cluster center. These completeness curves are used for completeness correction of stars in the rest of this paper. In addition, we also estimate the general completeness as function of magnitude by comparing the DECaLS catalog with the deep COSMOS catalog of \citet{2007ApJS..172..219L} regardless of the crowdedness. If this completeness is applied in the following analyses, it makes no big difference in the results.


\section{Determining Basic Properties of Palomar 5} \label{sec:property}
\subsection{Structure Parameters}
To estimate the structure parameters of Palomar 5 from the radial surface density profile, we first need to determine the central position of the cluster.  The Gaussian Kernel Density Estimation (KDE) technique\footnote{\url{https://docs.scipy.org/doc/scipy/reference/generated/scipy.stats.gaussian_kde.html}}
 is used to estimate the probability density function of the cluster spatial distribution in a nonparametric way. A two-dimensional KDE density map is constructed and the location with the maximum density is considered to be the center of Palomar 5. The central coordinate is determined as ($\alpha=229.\degree019$, $\delta=-0.\degree122$).

\begin{figure}[ht!]
\centering
\includegraphics[width=\columnwidth]{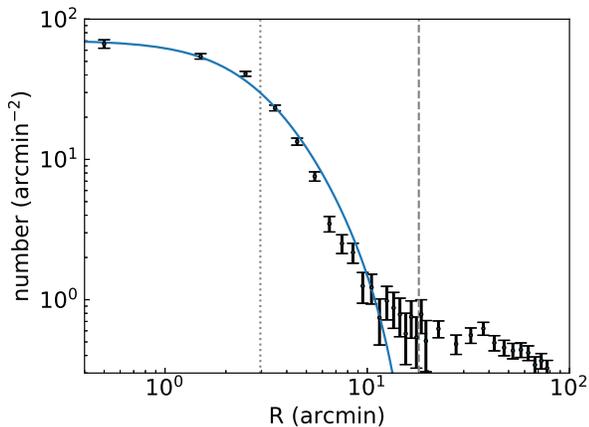}
\caption{Stellar surface density profile of Palomar 5. The average background density of field stars (about 4.61 arcmin$^{-2}$) is subtracted from the density profile. The black point is the surface density at a specified distance and the error bar presents the corresponding 1$\sigma$ Poisson error. The solid line shows the best-fit King model profile. The dotted line represents the core radius and dashed line represents the tidal radius. \label{fig:king}}
\end{figure}

The radial surface density profile of Palomar 5 is calculated according to the center as determined above.  Adopting the procedure described in \citet{1999ApJ...522..983F}, we calculate the density profile by using conventional star counts in a series  of apertures. The sample of stars is divided into different concentric annuli, and the stellar density in each annulus is computed as the ratio between the number of stars and the area of the annulus. The average background density of field stars is computed in a specified region far away from the cluster and is subtracted from the stellar density. The density is also corrected for the completeness as mentioned previously. The width of annuli is set to 1{\arcmin} at the distance of $R<20\arcmin$ and set to 5{\arcmin} outside of $R=20\arcmin$. The radial surface density profile of Palomar 5 is shown in Figure \ref{fig:king}.

The King model \citep{1962AJ.....67..471K} is widely used as a standard model to analyze the dynamics of the star cluster. The cluster structure parameters can be determined by fitting a \citet{1962AJ.....67..471K} model to the radial surface density profile. The empirical King model is described as
\begin{equation}
f(R)=k\left(\frac{1}{\sqrt{1+\left(R / r_{c}\right)^{2}}}-\frac{1}{\sqrt{1+\left(r_{t} / r_{c}\right)^{2}}}\right)^{2},
\label{equ:king}
\end{equation}
where $k$ is the scale coefficient, $r_c$ is the cluster core radius, $r_t$ is the cluster tidal radius, and $R$ represents the distance from the cluster center. The core radius is defined as the distance from the cluster center, where the stellar density falls to half of cluster central density. The tidal radius is at the distance where the stellar density falls to zero. Once member stars of a cluster exceed the tidal radius, they will be stripped by tidal forces of the Milky Way. The logarithmic ratio of the tidal radius to core radius ($c= \log_{10}(r_t/r_c)$) gives the central concentration parameter.  We use the above empirical King model as described in Equation (\ref{equ:king}) to fit the radial surface density profile and derive those structure parameters. The curve in Figure \ref{fig:king}  shows the best-fit King model profile. We obtain the core radius of $r_c=2.^{\prime}96\pm0.^{\prime}11$, tidal radius of $r_t=17.^{\prime}99\pm1.^{\prime}49$, concentration of $c=0.78\pm0.04$. The concentration of 0.78 indicates that Palomar 5 is a relaxed star cluster in virial equilibrium \citep{2011ApJ...739...15S}. At the same time, we can see from Figure \ref{fig:king} that the wing of the profile at $R>13\arcmin$ is located above the King profile, showing the low-density tidal tails. The best-fit structure parameters are also listed in Table \ref{tab:stlit}. 

We compare our estimates of structure parameters with previous determinations from literature in Table \ref{tab:stlit}. Our core or tidal radius is similar to those values in most previous studies \citep{1977AJ.....82..459S,1979ARAA..17..241H,1995AJ....109..218T,2002AJ....124.1497O}. In all previous studies, the photometric data are from either SDSS or early photoplate/CCD images. The imaging depths are much shallower than ours, and none of them have corrected for the incompleteness of source detection. The incompleteness is more serious in the outer region of the cluster due to the mass  segregation, which might underestimate the tidal radius.  The core and tidal radii of \citet{2009RAA.....9.1131Z} are significantly smaller than ours and other measurements, probably because they lack the innermost surface density point in their radial profile that is key for the core radius estimation and imperfect King model fitting for the wing of the profile. \citet{2017ApJ...842..120I} estimated a tidal radius of $r_t=21^{\prime}_{\cdot}2$ that is substantially larger than ours. They used the photometric data from the CFTH catalog with depths and completenesses similar to the DECaLS data.  However, the structure parameters in \citet{2017ApJ...842..120I} are derived by fitting the radial profile with the \citet{1966AJ.....71...64K} model, which includes four free parameters. As shown in Figure 10 of  \citet{2017ApJ...842..120I}, the tidal radius parameter is obviously coupled with other parameters, resulting in a large uncertainty.  

\begin{deluxetable}{ccc}
\tablecaption{The structure parameters of Palomar 5 from This Work and Previous Literature\label{tab:stlit}}
\tablewidth{0pt}
\tablehead{
\colhead{Core radius (\arcmin)} & \colhead{Tidal radius(\arcmin)} & \colhead{References}
}
\startdata
$2^{\prime}_{\cdot}96\pm0^{\prime}_{\cdot}11$ & $17^{\prime}_{\cdot}99\pm1^{\prime}_{\cdot}49$ & (1) \\
$3^{\prime}_{\cdot}6\pm0^{\prime}_{\cdot}2$ & $16^{\prime}_{\cdot}1\pm0^{\prime}_{\cdot}8$ & (2) \\
$1^{\prime}_{\cdot}68$ & $14^{\prime}_{\cdot}28$ & (3) \\
$2^{\prime}_{\cdot}9$ & $15^{\prime}_{\cdot}9$ & (4) \\
& $21^{\prime}_{\cdot}2$ & (5) \\
& $18^{\prime}$ & (6) \\
& $18^{\prime}_{\cdot}2$ & (7) \\
\enddata
\begin{tablenotes}
\footnotesize
\item \textbf{References.} (1) This work, (2) \citet{2002AJ....124.1497O}, (3) \citet{2009RAA.....9.1131Z}, (4) \citet{1995AJ....109..218T}, (5) \citet{2017ApJ...842..120I}, (6) \citet{1977AJ.....82..459S}, (7) \citet{1979ARAA..17..241H}.
\end{tablenotes}
\end{deluxetable}

\subsection{Stellar population parameters}
\subsubsection{Bayesian Analysis with BASE-9}
The stellar population properties and the distance of a cluster can be well determined by comparing observed CMDs  to theoretical stellar isochrones. The traditional method to estimate these parameters is fitting a single CMD with models by eye. Here we adopt a Bayesian analysis method, BASE-9\footnote{\url{https://github.com/argiopetech/base/releases)}} \citep{2014arXiv1411.3786V}, to estimate the distance modulus and stellar population parameters for Palomar 5.  BASE-9 is developed to recover the cluster and stellar parameters from multiwavelength photometric data. It implements a Markov Chain Monte Carlo (MCMC) algorithm to derive the posterior probability distributions for up to six cluster and three stellar properties. Here we care about the cluster parameters including the age, metallicity, line-of-sight dust absorption, and distance modulus. Compared with the classical method of isochrone fitting that is only applied to a single CMD, BASE-9 allows simultaneously fitting a set of parameters and make full use of the multicolor photometric data. It is expected that BASE-9 would generate more reliable results.  

BASE-9 provides a choice of a large library of stellar evolution models for stars at the main-sequence and red- giant branch stages and for white dwarfs \citep{1995PASP..107.1047B, 1998MNRAS.296..206A, 1999ApJ...525..482M, 2000A&AS..141..371G, 2008ApJS..178...89D, 2001ApJS..136..417Y, 2010ApJ...717..183R}. We select PARSEC models as provided in BASE-9, which covers a range of logarithmic age (log(age)) from 7.42 to 10.13 with a step of $\sim0.01$ dex and and six [Fe/H]s from -2 to 0.5 dex with a step of 0.5 dex. The grid of PARSEC isochrones only includes the photometric bands of Gaia, SDSS, and 2MASS, so that we need to convert our photometric data to the SDSS photometric system. We derive the system transformation equations for converting the DECaLS magnitudes to the SDSS ones as follows: 
\begin{eqnarray}
g_\mathrm{SDSS} &=& g_\mathrm{DECaLS}+0.1105(g-r)_\mathrm{DECaLS}+ 0.0092 ,\\
r_\mathrm{SDSS} &=& r_\mathrm{DECaLS}+ 0.1028(r-z)_\mathrm{DECaLS}+0.0434,\\
z_\mathrm{SDSS} &=& z_\mathrm{DECaLS}+0.0328(r-z)_\mathrm{DECaLS}-0.0129.
\end{eqnarray}
These transformation equations are determined based on stars with correction for Galactic extinction and valid for stars with $-0.5<(g-r)_\mathrm{DECaLS}<1.7$ or $-0.5<(r-z)_\mathrm{DECaLS}<2.7$.  We only consider single stellar population models so as to reduce the parameter degeneracy, i.e., the binaries are not included in the models. It is believed that GCs have low binary fractions and  binary stars would not have a significant effect on the determination of other properties \citep{2017MNRAS.468.1038W}.

\begin{figure*}[ht!]
\centering
\includegraphics[width=0.9\textwidth]{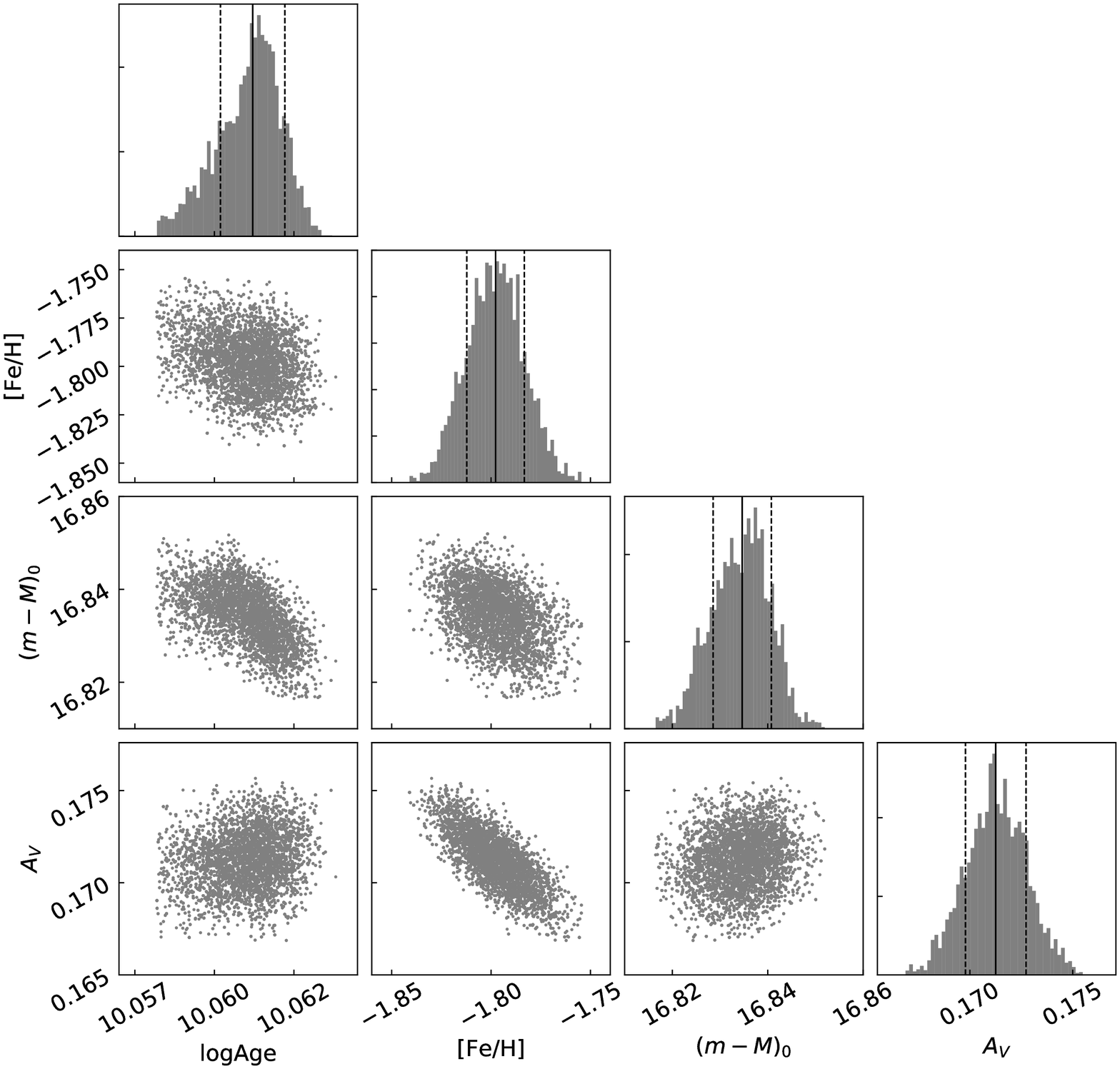}
\caption{Posterior distributions and correlations for fundamental parameters of Palomar 5 by using BASE-9, including log(age), [Fe/H], $(m-M)_0$, and $A_V$. The solid and dashed lines indicate the median value and $1\sigma$ error of each parameter. \label{fig:base9}}
\end{figure*}

\begin{figure*}[ht!]
\centering
\includegraphics[width=0.9\textwidth]{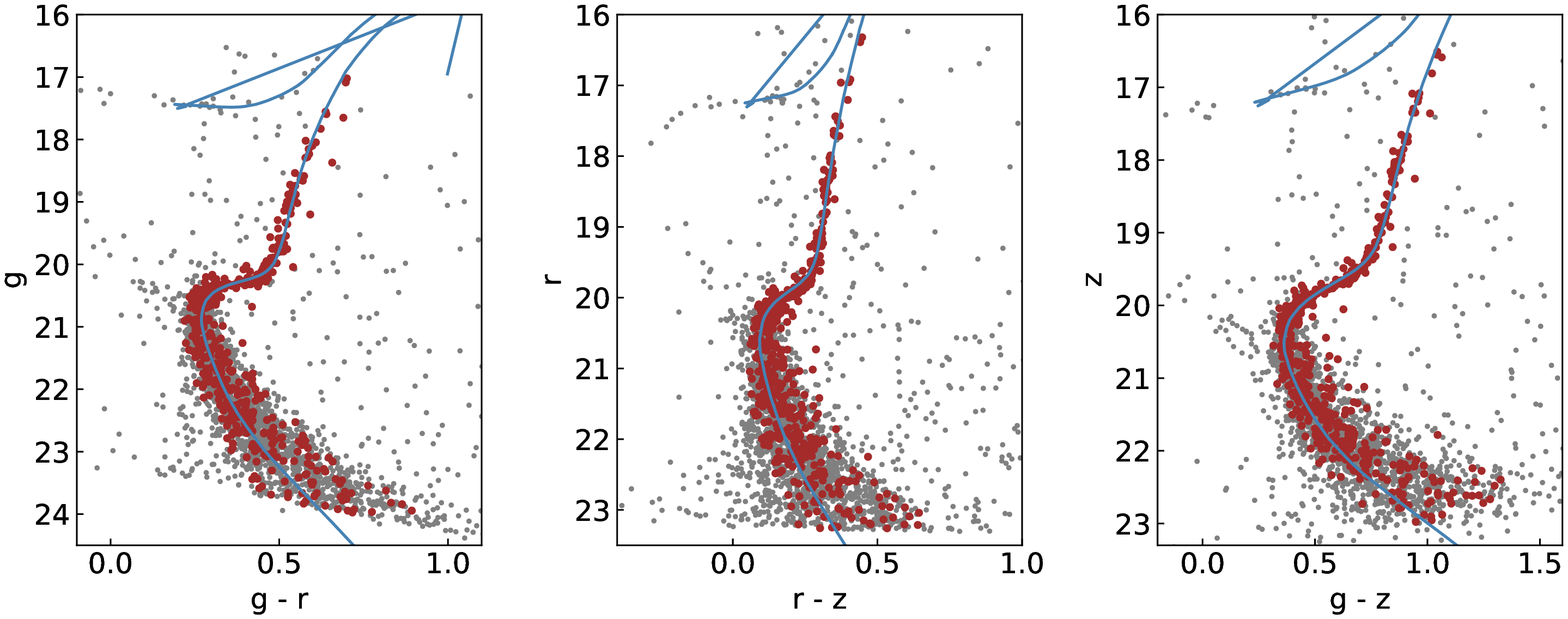}
\caption{The observed CMDs overplotted with the best-fit isochrones. The gray data points are the stars with $R<6\arcmin$. The red points are selected stars used for deriving the stellar population properties by BASE-9. The solid curves are Padova theoretical isochrones corresponding to the best-fit parameters.  \label{fig:basecmd}}
\end{figure*}

The prior distributions of the parameters are needed in BASE-9. We assume a uniform prior on age and Gaussian priors on metallicity ([Fe/H]),  intrinsic extinction ($A_V$) and distance modulus \citep[$(m-M)_0$;][]{2016MNRAS.463.3768W,2017MNRAS.468.1038W}. The Gaussian means are -1.4 dex, 0.09 mag, and 16.8 mag for [Fe/H], $A_V$ and $(m-M)_0$, respectively, which are taken as the average of the literature values (see Table \ref{tab:plit}).  The corresponding standard deviations are set to be 0.5 dex for [Fe/H], 1 mag for $A_V$ and 2 mag for $(m-M)_0$. These dispersions are conservatively set so as to decrease the impact of prior information on posterior probability distributions of the parameters \citep{2016MNRAS.463.3768W}.  We only select stars located close to the cluster center ($R<6\arcmin$, which is about 2 times the core radius) in order to reduce the contamination from field stars. Members stars are selected as those points within two manually designed polygons in the CMDs of $g$ vs. $g-r$ and $r$ vs. $r-z$, which embrace the main-sequence, subgiant and red-giant branches of the cluster. Following \citet{2016ApJ...826...42W}, we randomly select the subsample of those member stars, making the equal number of stars above and below the main-sequence turnoff point at $g\sim20.7$. After implementing BASE-9, we obtain the posterior distributions of the above four cluster parameters from 10,000 MCMC samples, which are shown in Figure \ref{fig:base9}. The histogram in this figure shows the posterior probability distribution of each parameter. The scatter plot of Figure \ref{fig:base9} shows the parameter correlation in the solution space. From these plots, we can see that there is a slight parameter degeneracy between [Fe/H] and $A_V$.  The parameter values and corresponding uncertainties are estimated as the medians and standard deviations: log(age) =$10.061\pm0.001$ (age of $11.508\pm0.027$ Gyr), $(m-M)_0=16.835\pm0.006$ ($23.281\pm0.064$ kpc), $E(B-V)$=$0.0552\pm0.0005$ ($A_V=0.171\pm0.001$), and [Fe/H]=$-1.798\pm0.014$. The parameter error only reflects the internal uncertainty of BASE-9, and it should be underestimated \citep{2016MNRAS.463.3768W, 2016ApJ...826...42W, 2018A&A...615A..49C}. We plot the observed CMDs overplotted the best-fit isochrones in Figure \ref{fig:basecmd}. It can be seen in this figure that the isochrones match well with the photometric data.

\subsubsection{Comparison with the traditional CMD fitting method and literature results} \label{sec:sp_comp}
As a comparison of the above Bayesian analysis method, we estimate the same stellar population parameters by using a traditional CMD fitting method \citep[Automated Stellar Cluster Analysis package, ASteCA;][]{2015A&A...576A...6P}. Briefly, ASteCA builds a large set of synthetic CMDs, which are generated from theoretical isochrones for a series of fundamental parameters including age, metallicity, reddening, and distance modulus. The CMDs are resampled according to a specified initial mass function and a total stellar mass. The fraction of binary stars can also be considered. These CMDs are then perturbed by the observed magnitude completeness and error functions, which simulate the observational characteristics. The optimization algorithm yields the best-fit parameter solutions by matching the synthetic and observed CMDs. A detailed description of the method can be found in \citet{2015A&A...576A...6P}. To be consistent with BASE-9, the PARSEC isochrone models are adopted and binary stars are not included in the models. The prior settings for the parameters are as close as possible to those of BASE-9. The CMDs of $g$ vs. $g-r $, $r$ vs. $r-z$, and $z$ vs. $g-z$ are separately fitted by ASteCA, and the best-fit parameters are listed in Table \ref{tab:parcmd}. From this table, we can see that the parameters derived from different CMDs vary quite a bit. There is relatively strong parameter degeneracies among age, metallicity, and distance modulus. The weighted averages of the parameters through three CMD fittings are also presented in Table \ref{tab:parcmd}. The average reddening and metallicity derived by ASteCA are consistent with those derived by BASE-9, but ASteCA gives a slightly larger age and smaller distance modulus due to the parameter degeneracy. However, because BASE-9 utilizes multiwavelength data simultaneously, it is expected that the parameter derivation should be more reliable. 

\begin{deluxetable*}{cccccc}
\tablecaption{Stellar population parameters derived by ASteCA\label{tab:parcmd}}
\tablewidth{0pt}
\tablehead{
\colhead{CMD} & \colhead{Age (Gyr)} & \colhead{$(m-M)_0$} & \colhead{$E(B-V$)} & \colhead{[Fe/H]}
}
\startdata
$g$ versus $g-r$ & 11.668$\pm$0.081 & 16.700$\pm$0.016 & 0.058$\pm$0.002 & -1.501$\pm$0.036 \\
$r$ versus $r-z$ & 12.162$\pm$0.168 & 16.622$\pm$0.029 & 0.039$\pm$0.002 & -1.961$\pm$0.077 \\
$z$ versus $g-z$ & 12.647$\pm$0.116 & 16.625$\pm$0.021 & 0.053$\pm$0.002 & -1.961$\pm$0.104 \\
mean $\pm$ std & 12.159$\pm$0.400 & 16.649$\pm$0.036 & 0.050$\pm$0.008 & -1.808$\pm$0.217 \\
\enddata
\end{deluxetable*}

As a comparison, we also collect the stellar population parameters from previous studies in Table \ref{tab:plit}. Most stellar population parameters are determined through the classical isochrone fitting method, such as \citet{1986AJ.....91..842S}, \citet{2002AAS...201.0711M}, \citet{2004AA...422..205G}, \citet{2011ApJ...738...74D}, and \citet{2014AA...564A..18P}. Our age matches well with other estimates in the literature, except the one derived by \citet{1986AJ.....91..842S}, who got a much larger age of 14 Gyr. Most of the metallicity estimations are based on the spectral measurements, and these metallicities show some dispersion. In general, our metallicity is slightly smaller than the literature values. Because there are only several metallicities available in the isochrone models, the interpolation might induce a large uncertainty. Our dust reddening of $E(B-V)=0.0552$ is close to the average of $E(B-V)=0.08$ from Hubble Space Telescope (HST) photometry of \citet{2011ApJ...738...74D} and $E(B-V)=0.03$ from other studies. Our distance modulus is similar to that based on the isochrone fitting with the photometric data. The distance modulus derived from RR Lyrae is smaller than ours and other measurements, which should be a systematic difference due to the application of different methods.

\begin{deluxetable*}{cccccc}
\tablecaption{Stellar population parameters of Palomar 5 derived in This Work and Previous Literature\label{tab:plit}}
\tabletypesize{\tiny}
\tablewidth{0pt}
\tablehead{
\colhead{age (Gyr)} & \colhead{$(m-M)_0$} & \colhead{$E(B-V)$} & \colhead{[Fe/H]} & \colhead{Data/Method} & \colhead{References}
}
\startdata
11.508$\pm$0.027 &16.835$\pm$0.006& 0.0552$\pm$0.0005 & -1.798$\pm$0.014 &  DECaLS photometry, BASE-9 & (1) \\
 & 16.56 &   & -1.3  & Keck spectra, $(m-M)_0$ from RR Lyrae, [Fe/H] from spectra &   (2) \\
  & 16.86 &   &   & SDSS photometry, modeling the tidal steam   &    (3) \\
 &  &   & -1.35$\pm$0.06 & Subaru and Keck spectra, spectral measurement  &    (4) \\
12.0$\pm$1.0 & 16.86 & 0.08 & -1.4 & HST $VI$ photometry, isochrone fitting & (5) \\
11.2 &  &   &  & SDSS photometry, isochrone fitting &  (6) \\
14 & 16.9 &  0.03 & -1.41  & 4 m CTIO telescope $BV$ photometry, isochrone fitting & (7) \\
  &   &  & -1.4 & MMT spectra, spectral measurement & (8) \\
  &   &  & -1.56$\pm$0.02  & Keck spectra, spectral measurement &  (9) \\
  &  16.75 &  0.03 & -1.11 & 5 m Hale spectrometer,  $(m-M)_0$ from RR Lyrae, $E(B-V)$ and [Fe/H] from spectra, &  (10) \\
  & 16.82 &   &   & CFHT photometry, isochrone fitting & (11) \\
  & 16.80 &   &   & SDSS photometry, modeling the tidal steam & (12) \\
  & 16.66 & 0.03 &   & $UBVR$ photoplate photometry,  $(m-M)_0$ from RR Lyrae, $E(B-V)$ from ultraviolet excess & (13) \\
  &  &  & -1.48$\pm$0.10& AAT spectra, spectral measurement& (14) \\
  &   &  & -1.52$\pm$0.28 & Washington system photometry, color-color diagram & (15) \\
 11.48$\pm$1 &16.86 & & & HST $VI$ photometry, isochrones fitting &  (16) \\
\enddata
\begin{tablenotes}
\footnotesize
\item \textbf{References.} (1) This work, (2) \citet{2002AJ....123.1502S}, (3) \citet{2015ApJ...803...80K}, (4) \citet{2016ApJ...823..157I}, (5) \citet{2011ApJ...738...74D}, (6) \citet{2004AA...422..205G}, (7) \citet{1986AJ.....91..842S}, (8) \citet{1985ApJ...298..249S},  (9) \citet{2017AA...601A..41K}, (10) \citet{1978ApJ...225..357S}, (11) \citet{2014AA...564A..18P}, (12) \citet{2016ApJ...833...31B}, (13) \citet{1977AJ.....82..459S}, (14) \citet{2015MNRAS.446.3297K}, (15) \citet{1997PASP..109..799G}, and (16) \citet{2002AAS...201.0711M}.
\end{tablenotes}
\end{deluxetable*}

\section{Luminosity and mass functions} \label{sec:luminosity}
The luminosity function (LF) of a GC can reveal its dynamical evolution. Based on the data taken by the HST,  \citet{2001AJ....122.3231G} reported that Palomar 5 has a flat luminosity function at the faint magnitude end, which indicates a significant loss of low-mass stars. They suggested that Palomar 5 has undergone strong tidal stripping. \citet{2004AJ....128.2274K} had a more in-depth study on Palomar 5 using the HST data and drew similar conclusions. They investigated the luminosity functions in different regions of Palomar 5 and find that the LF of the cluster center is flatter than that of the outer regions and tidal tails, showing clear evidence of mass segregation and tidal stripping. 

\begin{figure*}[ht!]
\centering
\includegraphics[width=1\textwidth]{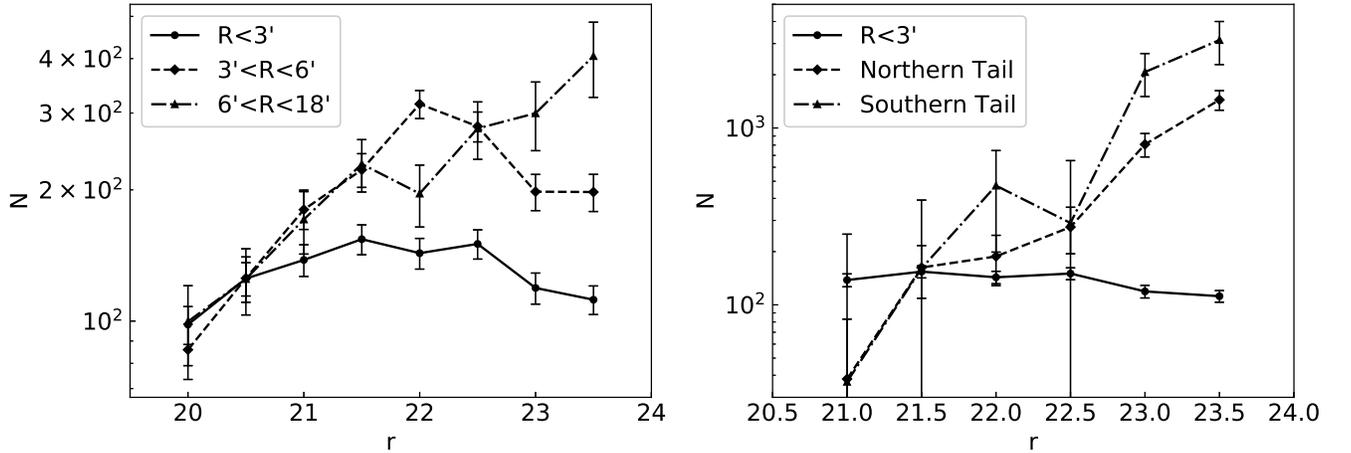}
\caption{Left: luminosity functions in three different regions: $R<3\arcmin$, $3\arcmin<R<6\arcmin$, and $6\arcmin<R<18\arcmin$. Right: luminosity functions in the cluster center region ($R<3\arcmin$) and in the regions of the tidal tails. Error bars present the Poisson errors from number counts.  \label{fig:lf}}
\end{figure*}

We construct the main-sequence luminosity functions for Palomar 5 to check its mass loss. The steps for eliminating the contamination from field stars are followed from \citet{2004AJ....128.2274K}. The $g$ vs. $g-r$ CMD of stars within the core radius of Palomar 5 is plotted, and an artificial envelope around the cluster main sequence is drawn to separate the cluster members from the field stars. The stars located in the envelope are considered as the cluster member candidates. The contamination of field stars is estimated by counting the number density of stars within the same envelope in an area of 0.5$\times$0.5 deg$^2$ far away from the center of Palomar 5. The main-sequence luminosity function is calculated by subtracting number density of field stars from that of cluster candidates in a series of $r$-band magnitude bins (at an interval of 0.5 mag).  The luminosity function is corrected for incompleteness with the completeness curve as described in Section \ref{sec:data}. The left panel of Figure \ref{fig:lf} shows the luminosity functions for different regions of Palomar 5 at different distances from the center ($R<3\arcmin$, $3\arcmin < R < 6\arcmin$ and $6\arcmin < R <18\arcmin$). These regions are roughly determined according to the core radius of $\sim$3{\arcmin} and tidal radius of 18{\arcmin} and the principle for making a similar number of member stars in each region. The LF curves are shifted to the bright end of the central LF in order to compare the steepness of LFs from different regions. We can see in this figure that the LF becomes steeper with increasing distance. The central LF is deficient in low-mass stars, and the outermost LF presents much more low-mass stars. The right panel of Figure \ref{fig:lf} shows two additional LFs for the northern and southern tidal tails. Here, the regions for tidal tails are similar to the ones defined in \citet{2004AJ....128.2274K}. Note that the LFs at the bright end ($r<21$) are not displayed due to a lack of bright or high-mass stars in the tails. The LFs of the tidal tails present more abundant low-mass stars than the central LF, indicating that the tails are enhanced by low-mass stars. The plots in Figure \ref{fig:lf} suggest that Palomar 5 shows prominent characteristics of mass segregation and substantial mass loss of low-mass stars into the tidal tails, which are consistent with previous studies.  

\begin{figure}[ht!]
\centering
\includegraphics[width=\columnwidth]{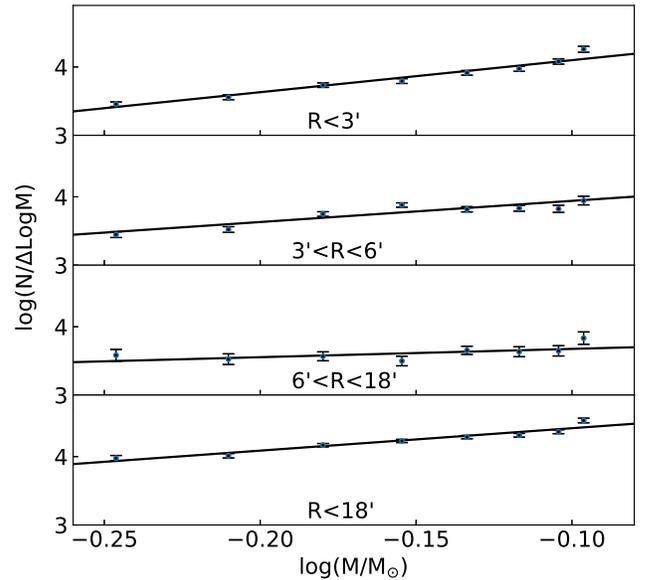}
\caption{Mass functions of Palomar 5 in different regions ($R<3\arcmin$, $3\arcmin < R < 6\arcmin$, $6\arcmin < R < 18\arcmin$, and $R<18\arcmin$ from top to bottom panels) with the corrections of completeness and contamination from field stars. The error bars are 1$\sigma$ Poisson error derived from number counts. The solid lines show the best-fit power laws. \label{fig:mf}}
\end{figure}

The mass function (MF) of Palomar 5 is built through converting the luminosity function using the mass-luminosity relation (M/L relation). The M/L relation is obtained from a theoretical PARSEC isochrone with log(age)=10.06 and [Fe/H]=-1.79 derived in this work. The MFs at different distances from the center of Palomar 5 are presented in Figure \ref{fig:mf}. The MF is usually defined as a power law of $\varphi(\log(m)) \propto m^{\alpha}$.  The slope $\alpha$  in the logarithmic space is derived through a linear least-squares fitting. The MF slopes at different distances are derived as below:
\begin{eqnarray}
\alpha&=&4.68\pm0.43, R<3\arcmin, \\
\alpha&=&3.10\pm0.65, 3\arcmin<R<6\arcmin,  \\
\alpha&=&1.21\pm0.60, 6\arcmin<R<18\arcmin,  \\
\alpha&=&3.28\pm0.39, R<18\arcmin.  
\end{eqnarray}
All the slopes are positive, suggesting that there is considerable mass loss of low-mass stars. As can be seen in Figure \ref{fig:mf}, the slope gradually decreases as the distance from the cluster center increases.  The variation of the MF slopes indicates that there are more massive stars in the inner region of the cluster, and more low-mass stars tend to be located in the outer region, which also shows the mass segregation in this GC. The mass segregation is considered to have resulted from the dynamical evolution: two-body relaxation dominates the evolution of a GC during the lifetime of the GC. \citet{2004AJ....128.2274K} suggested that the external effect like the Galactic tidal shock might also contribute to the mass segregation of Palomar 5 in addition to internal relaxation dynamics. From the mass function of Palomar 5, we derive the total cluster mass of 3825 $M_{\odot}$ within the tidal radius by integrating the mass function to a low-mass limit of 0.1 $M_\odot$. Similarly, we also obtain the total number of stars within the tidal radius as $N=8302$ . The relaxation time $t_{\text {relax }}$ is calculated as $t_{\text {relax}} \sim t_{\text {cross}} \frac{0.1N}{\mathrm{ln}N}$, where $t_{\mathrm{cross}}=D / \sigma_{v}$, $D$ is the cluster radius, $\sigma_{v}$ is the velocity dispersion, and $N$ represents the total number of stars \citep{2008gady.book.....B}. We adopt $\sigma_{v} = 1.1$ km s$^{-1}$ \citep{2002AJ....124.1497O} and obtain the relaxation time as $t_{\text {relax}}$ = 9.7 Gyr . The relaxation time is shorter than the age of Palomar 5, which indicates that Palomar 5 is a totally relaxed system. 

\section{Summary} \label{sec:summary}
As the first in a series based on the wide and deep DESI imaging data, this paper aimed to redetermine the fundamental properties of Palomar 5, which is an old GC in the Galactic halo and shows prominent tidal tails. The DESI imaging data are about 2 mag deeper than the SDSS data, which have been used in most previous studies on Palomar 5. The basic parameters derived in this paper are listed in Table \ref{tab:parameters}. The center of Palomar 5 is determined as the density peak using a two-dimensional KDE technique. The empirical King model is adopted to fit the radial surface density profile. Correspondingly, the tidal and core radii and concentration parameter are derived. We apply the Bayesian analysis method of BASE-9 to obtain the stellar population parameters including age, metallicity, dust reddening, and distance modulus. BASE-9 is competitive with traditional single-CMD fitting methods, because it utilizes multiwavelength photometric data simultaneously, which could relieve the parameter degeneracy. The fundamental parameters are also collected in this paper for comparisons from literature.

\begin{deluxetable}{cc}
\tablecaption{Fundamental Properties of Palomar 5 Derived in This Work.\label{tab:parameters}}
\tabletypesize{\small}
\tablewidth{0pt}
\tablehead{
\colhead{Parameter} & \colhead{Value}
}
\startdata
R.A. & 229.\degree019 \\
Decl. & -0.\degree122\ \\
Core radius & $2^{\prime}_{\cdot}96\pm0^{\prime}_{\cdot}11$\\
Tidal radius & $17^{\prime}_{\cdot}99\pm1^{\prime}_{\cdot}49$ \\
Concentration & 0.78$\pm$0.04 \\
Age & 11.508$\pm$0.027 Gyr \\
\text{[Fe/H]} & -1.798$\pm$0.014 \\
$E(B-V)$ & 0.0552$\pm$0.0005 \\
$(m-M)_0$ & 16.835$\pm$0.006 \\
Mass within tidal radius & 3825 $M_\odot$ \\
Relaxation time & 9.7 Gyr \\
\enddata
\end{deluxetable}
We also investigate the LF and MF of Palomar 5, with corrections for completeness and contamination from field stars. The LFs at different distances of the cluster and in the regions of the tidal tails are obtained. It is confirmed that there is an obvious mass segregation and mass loss of low-mass stars due to the strong tidal interaction with the Galactic potential. The mass functions are obtained through the M/L relation, which is based on the stellar population parameters derived in this paper. The power-law slope of the mass functions also suggests the signature of considerable mass loss of low-mass stars and mass segregation due to the dynamical two-body relaxation.\\

We thank the anonymous referee for thoughtful comments
and insightful suggestions that improve our paper greatly. This work is supported by Major Program of National Natural Science Foundation of China (No. 11890691, 11890693). This work is also supported by the National Natural Science Foundation of China (NSFC; grant Nos. 11733007, 11673027, 11873053, 12073035). This work is also supported by the National Key R\&D Program of China No. 2019YFA0405501.

The Legacy Surveys consist of three individual and complementary projects: the Dark Energy Camera Legacy Survey (DECaLS; NOAO Proposal ID \# 2014B-0404; PIs: David Schlegel and Arjun Dey), the Beijing-Arizona Sky Survey (BASS; NOAO Proposal ID \# 2015A-0801; PIs: Zhou Xu and Xiaohui Fan), and the Mayall $z$-band Legacy Survey (MzLS; NOAO Proposal ID \# 2016A-0453; PI: Arjun Dey). DECaLS, BASS, and MzLS together include data obtained, respectively, at the Blanco telescope, Cerro Tololo Inter-American Observatory, National Optical Astronomy Observatory (NOAO); the Bok telescope, Steward Observatory, University of Arizona; and the Mayall telescope, Kitt Peak National Observatory, NOAO. The Legacy Surveys project is honored to be permitted to conduct astronomical research on Iolkam Du'ag (Kitt Peak), a mountain with particular significance to the Tohono O'odham Nation.

NOAO is operated by the Association of Universities for Research in Astronomy (AURA) under a cooperative agreement with the National Science Foundation.

This project used data obtained with the Dark Energy Camera (DECam), which was constructed by the DES collaboration. Funding for the DES Projects has been provided by the US Department of Energy, the US National Science Foundation, the Ministry of Science and Education of Spain, the Science and Technology Facilities Council of the United Kingdom, the Higher Education Funding Council for England, the National Center for Supercomputing Applications at the University of Illinois at Urbana-Champaign, the Kavli Institute of Cosmological Physics at the University of Chicago, Center for Cosmology and Astro-Particle Physics at the Ohio State University, the Mitchell Institute for Fundamental Physics and Astronomy at Texas A\&M University, Financiadora de Estudos e Projetos, \text{Fundação} Carlos Chagas Filho de Amparo, Financiadora de Estudos e Projetos, \text{Fundação} Carlos Chagas Filho de Amparo \text{à} Pesquisa do Estado do Rio de Janeiro, Conselho Nacional de Desenvolvimento Cientifico e Tecnologico and the Ministerio da Ciencia, Tecnologia e Inovacao, the Deutsche Forschungsgemeinschaft and the Collaborating Institutions in the Dark Energy Survey. The Collaborating Institutions are Argonne National Laboratory, the University of California at Santa Cruz, the University of Cambridge, Centro de Investigaciones Energeticas, Medioambientales y Tecnologicas-Madrid, the University of Chicago, University College London, the DES-Brazil Consortium, the University of Edinburgh, the Eidgenossische Technische Hochschule (ETH) Zurich, Fermi National Accelerator Laboratory, the University of Illinois at Urbana-Champaign, the Institut de Ciencies de l'Espai (IEEC/CSIC), the Institut de Fisica d'Altes Energies, Lawrence Berkeley National Laboratory, the Ludwig-Maximilians Universitat Munchen, and the associated Excellence Cluster Universe, the University of Michigan, the National Optical Astronomy Observatory, the University of Nottingham, the Ohio State University, the University of Pennsylvania, the University of Portsmouth, SLAC National Accelerator Laboratory, Stanford University, the University of Sussex, and Texas A\&M University.

BASS is a key project of the Telescope Access Program (TAP), which has been funded by the National Astronomical Observatories of China, the Chinese Academy of Sciences (the Strategic Priority Research Program ``The Emergence of Cosmological Structures'' grant \# XDB09000000), and the Special Fund for Astronomy from the Ministry of Finance. The BASS is also supported by the External Cooperation Program of Chinese Academy of Sciences (grant \# 114A11KYSB20160057), and Chinese National Natural Science Foundation (grant \# 11433005).

The Legacy Survey team makes use of data products from the Near-Earth Object Wide-field Infrared Survey Explorer (NEOWISE), which is a project of the Jet Propulsion Laboratory/California Institute of Technology. NEOWISE is funded by the National Aeronautics and Space Administration.

The Legacy Surveys imaging of the DESI footprint is supported by the Director, Office of Science, Office of High Energy Physics of the US Department of Energy under contract No. DE-AC02-05CH1123, by the National Energy Research Scientific Computing Center, a DOE Office of Science User Facility under the same contract; and by the US National Science Foundation, Division of Astronomical Sciences under contract No. AST-0950945 to NOAO.


\bibliography{pal5}{}
\bibliographystyle{aasjournal}



\end{document}